# Machine Learning Based Probe Skew Correction for High-frequency BH Loop Measurements

Yakun Wang, Song Liu, Jun Wang, *Member*, IEEE, Binyu Cui, *Student Member*, IEEE, and Jingrong Yang, *Student Member*, IEEE

*Abstract*—Experimental characterization of magnetic components has grown to be increasingly important to understand and model their behaviours in high-frequency PWM converters. The BH loop measurement is the only available approach to separate the core loss as an electrical method, which, however, is susceptive to the probe phase skew. As an alternative to the regular de-skew approaches based on hardware, this work proposes a novel machine-learning-based method to identify and correct the probe skew, which builds on the newly discovered correlation between the skew and the shape/trajectory of the measured BH loop. A special technique is proposed to artificially generate skewed images from measured waveforms as augmented training sets. A machine learning pipeline is developed with the Convolutional Neural Network (CNN) to treat the problem as an image-based prediction task. The trained model has demonstrated a high accuracy and generalizability in identifying the skew value from a BH loop unseen by the model, which enables the compensation of the skew to yield the corrected core loss value and BH loop.

## I. INTRODUCTION

The characterization of core loss for high-frequency magnetic components (e.g. Fig. 1(a)) used in power conversion applications has been increasingly important to inform the design and virtual prototyping of power converters. The most common method for measuring high-frequency core loss is the two-winding BH loop measurement approach shown in Fig. 1(b), due to its capability of separating the core loss (i.e. excluding the copper loss) and suitability for rapid testing under high-frequency excitations without the need for reaching a thermal equilibrium, compared to the calorimetric approaches.

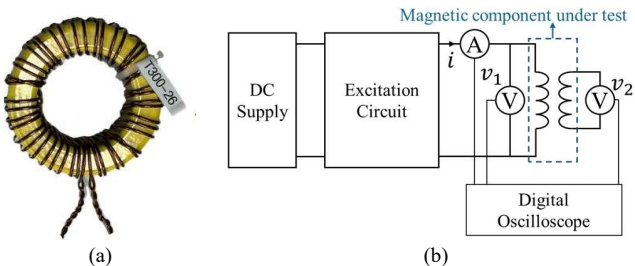

(a)          (b)

Fig. 1. (a) magnetic component for high-frequency power electronics applications (b) Two-winding BH loop measurement

However, the most challenging aspect of the two-winding method is the phase discrepancy error caused by the different propagation delays between the voltage and current probes, known as the skew - the probe skew can lead to a distorted BH loop resulting in up to ±200% of error in the measured core loss [1]. Conventionally, probe skew can only be calibrated through deskew tools, such as Keysight U1880A or Lecroy DCS025, which are designed to show a pair of voltage and current waveforms that are in phase to the probes. However, these tools cannot eliminate the skew due to physical imperfections on the hardware such as the inevitable parasitics in the circuits. Incorrectly measured magnetic core loss will misinform the design process of power electronic systems, which can lead to wrongly sized cooling components and potential failure of the converter systems in safety-critical applications, e.g. electric passenger cars.

Alternatively, methods have been proposed to alter the measurement and excitation circuits to minimize the impact of the probe skew. In [2], [3], the concept of reactive voltage cancellation is proposed to avoid capturing a pair of voltage and current waveforms that are almost in a 90° phase difference by placing reference components into the testing circuit. This concept's limitation is the requirement of the reference components which need to be precise matches to the device under test. [4] proposes to place a small capacitor in the testing circuit to induce a 'notch' in the measured voltage waveform which contains the information of probe skew to enable the correction. However, this approach is still challenging to practically implement as it requires a dedicated design of testing circuits with a swappable capacitor which is invasive for the main power circuit.

This article proposes a novel and timely solution to solve the probe skew problem in the two-winding BH loop measurement through a machine learning and data-driven approach, which does not require a local piece of hardware, i.e. a de-skew tool. The key concept is to establish a model that learns the mapping between the shape irregularity of the measured BH loop and the degree of skew, which is realized by a convolutional neural network (CNN) through a supervised learning process. For the first time, the correlation between the probe skew and the shape irregularity is revealed and evaluated in this work.

A training pipeline is developed and applied to the open-source MagNet dataset (https://github.com/minjiechen/magnetchallenge), which is a rich database of experimentally measured waveforms, along with a data augmentation technique to embed artificial skew values to create the training data. Fundamentally, the proposed machine learning approach treats the skew correction problem as an image recognition task. It does not suffer from the imperfection of a local de-skew tool, assuming the original training data is skew-free. The effectiveness of the proposed approach is validated on a testing set randomly selected from the dataset. Different to the existing literature that applies neural networks to model the characteristics of magnetic materials [13-15], this work focuses on the probe skew problem in the BH loop measurement that originates from the testing hardware, while it manifests as the shape irregularity in the BH loops.

This project has been supported by the Jean Golding Institute (JGI) for data science and data-intensive research at the University of Bristol. (Corresponding author: Jun Wang)

## II. PROBE SKEW IN BH LOOP MEASUREMENT

### A. Impact of probe skew on core loss measurement

To obtain the core loss, the two-winding method shown in Fig. 1 includes two steps: (1) measuring out the voltage $v_2$ and current $i$ waveforms and (2) converting the time-series of the electrical waveforms into flux density $B$ and field strength $H$, then, forming a BH loop where the area is the core loss.

$$B(t) = \frac{1}{N_2 A_e} \int_0^t v_2(t)\, dt \quad (B(0) = 0) \tag{1}$$

$$H(t) = N_1 \cdot i(t)/l_e \tag{2}$$

Where $N_1$ is the number of turns of the main winding of the inductor; $N_2$ is the number of turns of the flux-sensing winding; $A_e$ is the effective cross-section area of the core; $l_e$ is the effective magnetic path length of the core. In this process, the physical probe skew can be mimicked by shifting the measured waveforms horizontally along the time axisIn this work, the $H(t)$ waveform is the signal to be shifted to mimic the phase skew since it is directly proportional to the current waveform (2). This approach also echoes the fact that the voltage probe is typically the reference point in a probe calibration process. This technique enables the generation of data points with an artificial skew time.

The impact of the probe skew on measured core loss is shown in Fig. 2, treating the original data as the ground truth. It can be seen that the core loss deviation under one operating condition is in a linear relationship against the skew time in nanoseconds, which aligns with the findings reported in [1]. In practice, the delay of the current probe can range from a few to hundreds of nanoseconds. This linear relationship enables the correction of core loss once the skew time is known.

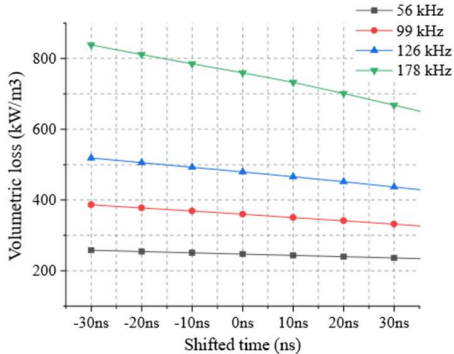

Fig. 2. 3C90 core loss ($\Delta B$ = 176 mT) with imposed skew time

### B. Shape irregularity in BH loops due to probe skew.

To correct the probe skew, the conventional approach is to supply a voltage (typically with sharp rising/falling edges) on a non-inductive resistor to generate a pair of voltage/current waves that are physically in phase so that the skew caused by the probe can be identified on an oscilloscope, which is the principle of the de-skew tools. As a fundamentally different approach, this work proposes to (1) convert the voltage/current waveform measured from a magnetic component into the BH domain and (2) identify the shape irregularity in the BH loops as a feature that links to the probe skew. Fig. 3 shows a BH loop measured experimentally through a Triple Pulse Test (TPT) procedure [5] on a ferrite core T184-26. Fig. 3(a) is the BH trajectory treated as the ground truth with the probes calibrated through a de-skew tool U1880A. Fig. 3(b) shows the BH loop once an artificial skew of 1.25 milliseconds is applied by shifting the data points of the current $I$ (equivalently the $H$) on the time axis. The distortion of the shape is visually noticeable in this case, and as a result, the core loss obtained from this loop sees a +123% difference compared to the original BH loop. This example demonstrates the dependency between the skew and the BH loop shape irregularity, which enables a pattern recognition technique based on machine learning to be developed for this problem.

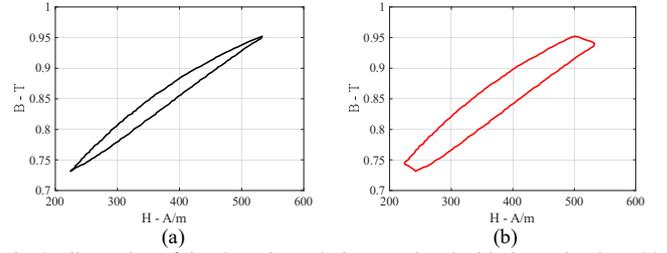

Fig. 3. Illustration of the shape irregularity associated with the probe skew (a) original BH loop (b) distorted BH loop with a skew of 1.25 millisecond

To further show the effect of skew, Fig. 4 illustrates a set of BH loops with both positive (delaying the current) and negative (delaying the voltage) artificial skews. As can be observed, with a negative skew, the BH loop area shrinks, and the trajectory shows incorrect crossovers. With a positive skew, which is the more common case where the current probe suffers more delays, the BH loop area expands, and the trajectory shows a 'flat top'. These observations correspond to the trend in loss value shown in Fig. 2 in both directions.

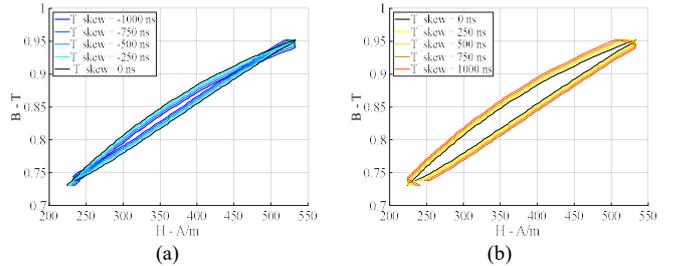

Fig. 4. Trend of BH loop shape distortion associated with various skew values (a) a negative skew (current leading) (b) a positive skew (current lagging). The colour gradient shows the trend of distortion.

## III. MACHINE LEARNING-BASED CORRECTION APPROACH

### A. Overall pipeline

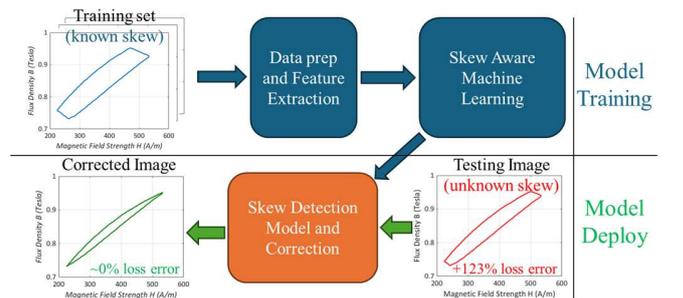

Fig. 5. Illustration of the proposed ML approach for skew detection

In recent years, CNN [6] [7] as a computer vision neural network architecture has been widely used in various sectors due to its capability to explore patterns in image data. For example, CNNs have been applied in healthcare applications for disease diagnosis [8], autonomous driving for object recognition [9], security for facial recognition [10], and agriculture for monitoring crop health and detecting plant diseases [11]. Inspired by the superb image recognition capability of CNNs, we propose a hyperparameter-free CNN approach for detecting the skew from high-frequency BH loop measurements. The high-level workflow of this approach is illustrated in Fig. 5. In the training stage, the training set is constructed using BH loops with known skew values. A supervised training process is then applied to fit a CNN-based skew detection model. As the outcome, the trained model can then be used to predict/detect the unknown skew (in degrees) associated with an unseen testing BH loop image. Using the detected skew value, the input BH loop and its associated core loss are then corrected/recalibrated to produce the un-skewed final output. In this work, the open-sourced MagNet database is used to construct training and testing sets [12].

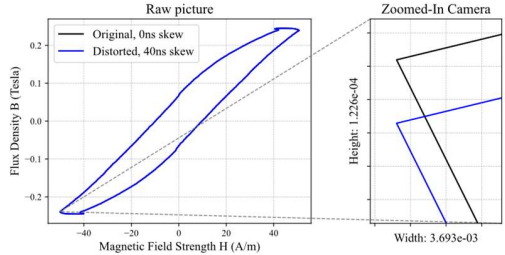

Fig. 6. Comparison of the shape irregularity with the original BH loop (black) and distorted BH loop with a skew of 40 ns (blue) from 3C90

### B. Data preprocessing

The MagNet dataset contains the raw time-series data of B and H from different materials. They are structured as

$$\begin{cases} B = [B_{t_1} \cdots B_{t_N}] \\ H = [H_{t_1} \cdots H_{t_N}] \end{cases} \quad (3)$$

By sequentially connecting the B-H coordinates $(H_{t_n}, B_{t_n})$ according to their index $t_n$, we can transform the raw time-series data into a "loop-shaped curve". These curves are referred to as the BH loops. This work uses the BH loops as the inputs for a machine-learning model to predict the skew. The data from the MagNet database are extracted and preprocessed as:

1. Preparing the dataset: The data used in this work is from the open-source MagNet dataset with data filters (e.g. frequency range, DC-bias, waveform types etc.) applied where applicable. The dataset extracted is then randomly split to training and testing sets with an 8:2 ratio.

2. Waveform interpolation: The original dataset consists of $B$ and $H$ as time series, each of which is stored as a 1024-dimensional vector. While the raw data has only 1024 points per cycle, linear interpolation is performed to expand the data points by 1000 times more to achieve a higher resolution. This operation effectively achieves a resolution of $3.5 \times 10^{-4}$ degrees. Note the interpolation is essential to enable a higher resolution for the data augmentation operation in the training stage and the skew correction operation in the deployment stage, while it does not make any difference in the generated images for training, which undergoes a downsampling operation.

3. Data augmentation: To create the training set with artificial skew values, each BH loop is augmented by an artificially imposed skew, which is realized by shifting the indexes of the $H$ sequence. For example, if the original time indexes are $[t_1, t_2, t_3, \ldots]$, the shifted indexes are $[t_1 + \delta, t_2 + \delta, t_3 + \delta, \ldots]$ and $\delta$ is the artificial skew index. If the range of $\delta$ is set as [-20, +20], one BH loop from the database can be expanded to a set of 41 BH loops each with a known skew value, including the original loop which has a skew of zero degrees.

4. Zooming-in: It is observed that a small skew (e.g. by 0.01 degrees) will lead to a distortion that is almost visually unidentifiable (see Fig. 6), implying that it may be difficult for CNNs to capture the visual differences in the image. To address this, an additional zoom-in camera is set at the bottom left of a BH loop, centered at the coordinate with the minimum value of $H$. This zoomed-in image typically includes patterns which are more recognizable and captures fine details caused by a small skew better. The zoomed image is placed next to the original BH loop as a part of the input image (see Fig. 7(b).).

5. Extra Scale Information: The size of the BH loop varies considerably under different operating conditions (see Fig. 7. (a)), thus, an automatic scaling and centering function is applied to each BH loop to obtain a reasonably sized training image. To factor in this automated scaling, the minimum and maximum values of $B$ and $H$ are also included as a part of the input, to indicate the actual scale of the original image ensuring that this information is not lost during the training process. This information is also fed to the zoom-in operation to adaptively place the camera on the bottom-left of a BH loop with a reasonable size to capture the details.

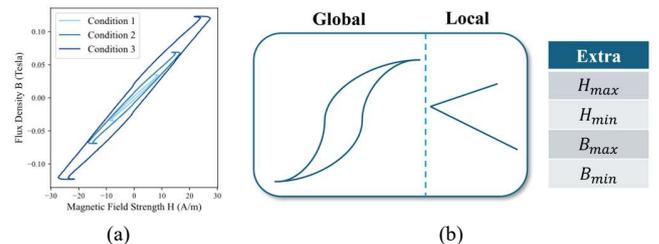

Fig. 7. Illustration of (a) different sizes of BH loops and (b) the input data to the neural network: the original BH loop, the zoomed-in view and four scalars

### C. The architecture of CNN

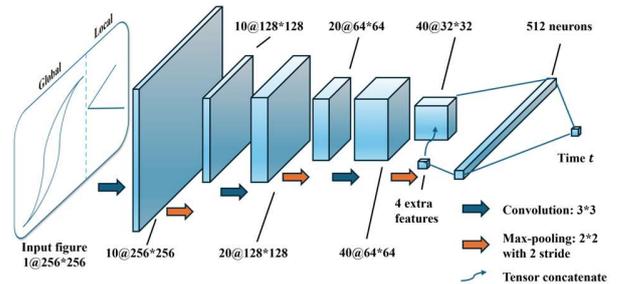

Fig. 8. Illustration of the CNN architecture

Fig. 8 summarizes the architecture of the proposed CNN model. The input picture is resized to 256×256 pixels (including both the original image and the zoomed-in view) to improve the computational efficiency. There are 8 layers in the CNN in total. Particularly, it comprises three convolutional layers, each followed by a max pooling layer. The first convolutional layer employs 10 filters with a kernel size of 3×3, stride of 1, and padding of 1. The second and third convolutional layers use 20 and 40 filters, respectively, maintaining the same kernel size and stride parameters. Following the last pooling layer, the output is flattened and then concatenated with a tensor storing four additional scale information (see Fig. 7.(b)). This combined feature vector is processed by two fully connected layers dedicated to the regression task. The first fully connected layer maps the input data to 512 hidden units, followed by a leaky ReLU activation function with a 0.01 negative slope. The second fully connected linear layer directly produces a single scalar output representing the predicted skew time.

Note there is little difference between concatenating the global and local view into a single image and feeding them separately into the model, since the convolutional kernel in the CNN scans across all pixels, extracting visual features from the entire image. As long as the local and global plots remain visually distinct, applying a CNN to a combined image should yield the same visual information as processing them separately. Moreover, merging the two plots into a single image reduces the input dimension and, consequently, the size of the neural network.

*D. Training and testing*

By applying the filters stated, two sub-datasets are extracted from the MagNet database, which are described below. Targeting representative power electronics applications, the following filters are applied in this instance to simplify the problem (1) triangular flux waveform (rectangular voltage) 2) temperature at 25 °C (3) DC bias at 1A/m. Two CNN models are trained for the 3C90 and N87 materials, which are the two most common and representative soft magnetic materials.

Table I Operating points used for training and testing

|  | $\Delta B$ (pk-pk, T) | Frequency (Hz) | Operating Points |
|---|---|---|---|
| 3C90 | [0.0215, 0.5524] | [56330, 446430] | 1197 |
| N87 | [0.0213, 0.5498] | [49970, 446430] | 1879 |

Following the data augmentation process, the total number of images is expanded to 1197×41 = 49077 for the 3C90 set and 77039 for the N87 set. The training and the testing operation points are randomly selected. The loss function is defined by

$$L(\theta) = \frac{1}{N}\sum_{i=1}^{N}|t_i - t'_i(\theta)|^2 \quad (4)$$

Where $N$ is the training sample size, $t_i$ is the target skew, artificially imposed during the data preprocessing and $t'_i(\theta)$ is the model prediction. The CNN is trained using ADAM optimizer by minimizing the loss $L(\theta)$ through backpropagation techniques with a learning rate set to $2.5\times10^{-3}$. Training of the CNN was performed in 50 epochs. For the 3C90 set, a batch size of 500 is used, with a training-to-testing ratio of 1000:197. For the N87 set, a batch size of 400 is used, with a training-to-testing ratio of 1600:279.

## IV. FUNCTIONALITY EVALUATION

To quantitively assess the performance of the proposed model, the relative error against the testing set for the two core materials 3C90 and N87 is shown below, containing 197×41=8077 and 11439 testing data points, respectively.

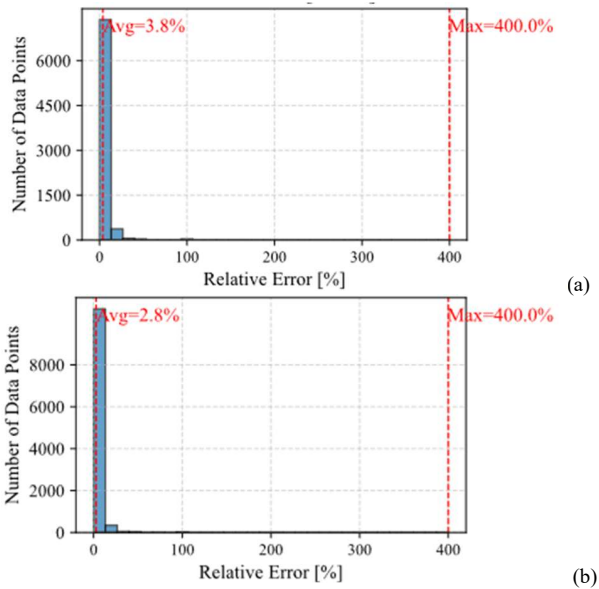

Fig. 9. Relative error distributions of identified skew (a) 3C90 (b) N87

Fig. 9 shows that the proposed method achieves remarkably low relative errors despite the outliers, which are 3.8% for 3C90 and 2.8% for N87 respectively, indicating the validity and the consistent performance of the CNN model across different materials and the effectiveness of the proposed approach. The validation on the MagNet database demonstrates that the proposed model can successfully capture the feature of the time skew from BH loops at a resolution of $3.5\times10^{-4}$ degrees (i.e. ~0.02 nanoseconds for the case with ~50 kHz).

Once the skew is predicted, the BH loop and the measured core loss can be corrected by shifting the *I* (or *H*) waveform to compensate for the skew, which is then followed by re-calculating the core loss. As a result of the interpolation, the skew correction operation can be performed at the same resolution of $3.5\times10^{-4}$ degrees. Fig. 10 provides a comprehensive evaluation of the relative error distributions of core loss before and after the skew correction, further demonstrating the validity of the proposed approach in the engineering context. The relative errors of the uncalibrated core loss points, indicated by red markers, show significant deviations across 5 frequencies arising from the imposed skew. After calibration, the relative errors of the new core loss, represented by blue markers, are significantly lower, with nearly all deviations shrinking to zero. Furthermore, Fig. 10(a) illustrates that the proposed CNN model effectively reduces the relative deviations for 3C90 across various frequency levels, verifying its robustness. Similarly, Fig. 9(b) confirms that the method is transferable to N87, achieving near-zero calibrated deviations across a diversified range of operation points. These results demonstrate the general applicability and accuracy of the proposed CNN-based correction method in de-skewing the high-frequency BH loop measurements across different materials and operating points.

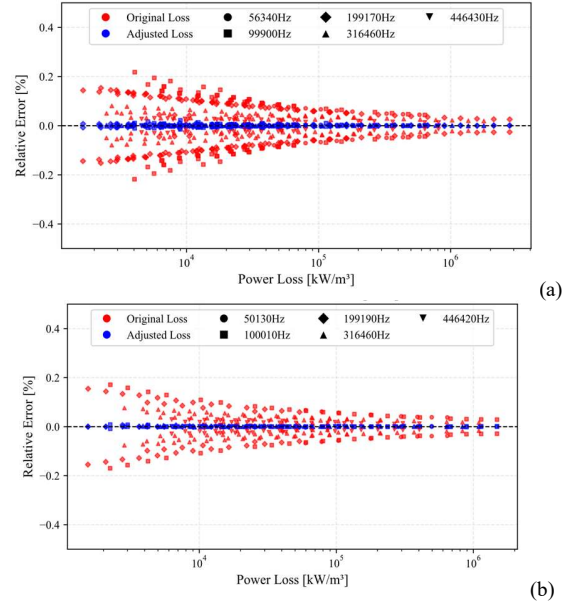

Fig. 10. Relative deviation of core loss (a) 3C90 (b) N87

## V. EVALUATION OF MODEL GENERALIZABILITY AND PROCESS ADAPTIVITY

To further evaluate the generality of the generated models and the adaptability of the proposed workflow, several comparative case studies are designed and conducted in this section. Note all the testing data used are waveforms/images that the models have never seen during the training process.

*A. Case 1 – Waveform challenge*

To investigate the impact of the training data composition regarding types of waveform shapes (e.g. sinusoidal, rectangular or trapezoidal), two cases are considered by adjusting the filters when the training and testing data is prepared. As illustrated in Fig. 11, this case study focuses on the data input of the process.

- **Case 1A** – only rectangular waveforms are selected as the baseline, which is the case in Section IV.
- **Case 1B** – all types of waveform shapes are considered with a balanced distribution, for which the dataset is kept as is from the MagNet database.

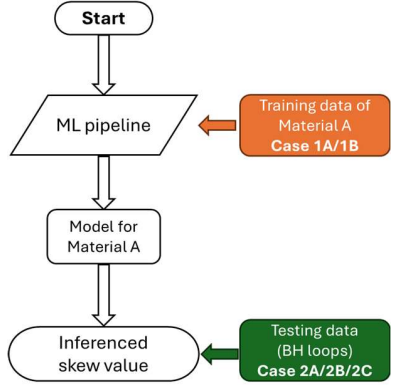

Fig. 11. Illustration of workflow for Case 1A/1B and Case 2A/2B/2C

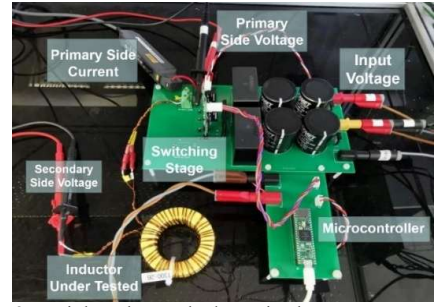

Fig. 12. Setup 2 – Triple Pulst Testbed at Bristol

The results are listed in Table II. It can be seen that, when all waveform types are included in the training and testing data, the process can still generate a model with high accuracy, though the average relative error rises to 14% from 3.8%. This case study confirms the proposed workflow is mostly effective for triangular $B$ waveforms. When it is applied to a mix of all waveform types, the task becomes more challenging.

Table II Results of Case 1 Evaluation

|  | Note | Average relative error of skew value |
|---|---|---|
| Case 1A | 3C90, 1197 operating points, squarewave only | 3.8% |
| Case 1B | 3C90, 2800 operating points, all waveform types, model retrained | 14.5% |

*B. Case 2 – Model generalizability*

Considering various application scenarios, the following cases are evaluated without altering the training workflow. This part assumes that a model has already been trained for Material A and it is directly deployed on various categories of testing data.

- **Case 2A** – the testing data is of the same material, Material A and from the experimental setup/context (Setup 1), which is the baseline that is described in Section IV.
- **Case 2B** – the testing data is of the same magnetic material (Material A) but from a different source, i.e. the waveforms are measured from a different experimental setup/circuit (Setup 2). This case emulates a practical application scenario of applying the model trained from one database (MagNet) to predict the skew in a given BH loop measured in a different setup by a user. For this instance, the testing data is measured locally from a Triple Pulse Testbed [5], of which a photo is shown in Fig. 12.
- Case 2C – the testing data is of a different (new) material, Material B, while the experimental setup remains the same (Setup 1). In this case, the data of a different material is pulled from the MagNet dataset as the testing data. This is to validate the general representativeness of the model trained from one soft magnetic material but used on a different one.

The results of this comparative study are shown below

Table III Results of Case 2 Evaluation

|  | Notes | Average relative error of skew value |
|---|---|---|
| Case 2A | 3C90, baseline, same as Section IV | 3.8% |
| Case 2B | 3C90 core, 5 sets of data measured from TPT | 21.4% |
|  | N87 core, 5 sets of data measured from TPT | 19.7% |
| Case 2C | 3C90 model used on N87 testing data (1879 operating points) | 13.7% |

The result of Case 2B shows that the trained model can still identify the skew in a given BH loop with a relatively high accuracy (1) measured from a different setup (2) featuring an operating point that is unseen by the model. A different measurement setup leads to minor discrepancies in the details of the waveform, e.g. the ringings in the voltage and current waveforms. This case shows that the model can moderately ignore the subtle/unimportant differences caused by the testing setup and identify the correct skew value. However, the results show a wider spread of cases with higher errors, indicating that the difference in testing setup has an unneglectable impact on the accuracy of the proposed approach. This challenge can be addressed if a large training dataset is available containing a mix of measurements from various testing setups.

The result of Case 2C demonstrates that the tested model is generalizable when it is directly used on a similar but different material (3C90 to N87), although the accuracy drops noticeably, e.g. with the average relative error increased from 3.8% to 13%. Due to limited data availability, this case is not tested on other core materials from a different family (e.g. nanocrystalline instead of ferrite). Overall, the recommendation is one model per material to achieve high accuracy, in contrast to Case 2C.

*C. Case 3 – Fine-tuning on a new material*

Although the generalizability of the model trained from one core material's data is proven acceptable, it is still preferred to incorporate the data from a new material to generate a model dedicated to it. Considering practical scenarios where a relatively small dataset is available for a new material, a refined workflow is investigated in this case which adds a fine-tuning stage to factor in the new data in the training. Two cases are considered as described for which the workflow is illustrated in Fig. 13.

- **Case 3A** – rich data for new material. This represents the case where a fully automated testing setup is available.
- **Case 3B** – small dataset for new material. This is a few-shot fine-tuning scenario which may be a result of a limited automation level of the testing setup.

The result of this comparative study is shown in Table IV. As can be seen, the achieved relative errors in Case 3A and 3B are both superior to Case 2C in Table III, which means the fine-tuning stage can enhance the model performance on a new material, since the model incorporates new data. In the few-shot scenario, the proposed workflow can still achieve a 10.4% average relative error in the tested case.

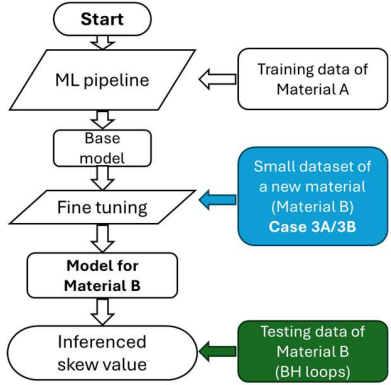

Fig. 13. Revised workflow with a fine-tuning stage for Case 3A/3B

Table IV Results of Case 3 Evaluation

|  | Notes | Average relative error of skew value |
|---|---|---|
| Case 3A | 3C90 as base model (~1197 points) fine-tuned with rich dataset of 3C94 (~2200 points) | 6.2% |
| Case 3B | 3C90 as base model (~1197 points) fine-tuned with a **small** (~200 points) data of 3C94 | 10.4% |

*D. Summary*

The comparative case studies have confirmed the excellent generalizability of the trained models considering the variations in waveform shapes, core materials and measurement setup. The evaluation results demonstrate the practical value of these models and the workflow when deployed in real-world applications. Following the proposed basic workflow, one skew-aware CNN model can be trained for one magnetic core material. If there is a need to adapt the existing model to a new material, the fine-tuning operation can be performed, even if there is very limited new data, to yield a more accurate model dedicated to the new material.

Note the presented models are trained on the MagNet dataset, treated as the ground truth despite potential skew. The proposed approach should be considered as a tool for aligning skew against a standard baseline rather than eliminating it. While the absolute skew cannot be fully determined, the relative skew between the testing data and the baseline dataset is identified in the proposed approach. The presented method can be applied to a skew-free dataset if one becomes available in the future.

Regarding the training data, ideally, it should contain a diversified mix of various testing conditions (e.g., frequency, amplitude or saturation), so that the trained model will then be effective for any conditions within the trained boundaries – this ensures that the proposed model is solving an interpolation problem rather than an extrapolation problem.

## VI. Conclusion

This work proposes a novel method to identify and compensate the probe skew in a measured BH loop to correct the core loss data for tested magnetics exposed in PWM waveforms. For the first time, the correlation between the probe skew and the shape irregularity is revealed and evaluated. A vision-based shape-aware calibration algorithm is then developed to capture the complex patterns embedded in the shape irregularity of BH loops. It can accurately identify prob skews in nanoseconds using these complex correlations. The algorithm is trained on the open-source MagNet database, which is an experimentally measured dataset. The proposed ML-based method is evaluated against testing sets on common core materials 3C90 and N87. The evaluations have demonstrated that the proposed approach is effective, accurate, and generalizable. Further case studies also show that the proposed models are generally effective for various types of waveforms, core materials and different setups. The models can also be adapted to a new material with limited data available by introducing a fine-tuning step in the training pipeline. The generalizability of the models can be further improved when a more comprehensive and diversified dataset becomes available. This approach can serve as a practical software alternative to the hardware-based de-skew tools for high-accuracy BH loop measurements.

References

[1] N. Rasekh, J. Wang, and X. Yuan, "A new method for offline compensation of phase discrepancy in measuring the core loss with rectangular voltage," IEEE Open J. Ind. Electron. Soc., vol. 2, no. April, pp. 302–314, Apr. 2021.
[2] M. Mu, Q. Li, D. J. Gilham, F. C. Lee, and K. D. T. T. Ngo, "New core loss measurement method for high-frequency magnetic materials," IEEE Trans. Power Electron., vol. 29, no. 8, pp. 4374–4381, 2014.
[3] D. Hou, M. Mu, F. C. Lee, and Q. Li, "New high-frequency core loss measurement method with partial cancellation concept," IEEE Trans. Power Electron., vol. 32, no. 4, pp. 2987–2994, 2017.
[4] L. Yi, M. McTigue, D. Gines, B. Doerr, and J. Moon, "Minimally Invasive Direct In-Situ Magnetic Loss Measurement in Power Electronic Circuits," IEEE Trans. Power Electron., vol. 38, no. 11, pp. 14334–14344, Nov. 2023.
[5] J. Wang, X. Yuan, and N. Rasekh, "Triple Pulse Test (TPT) for characterizing power loss in magnetic components in analogous to Double Pulse Test (DPT) for power electronics devices," in IEEE Proc. Annual Conference of the IEEE Industrial Electronics Society (IECON), 2020.
[6] Y. LeCun, K. Kavukcuoglu, and C. Farabet, "Convolutional networks and applications in vision," in Proceedings of 2010 IEEE International Symposium on Circuits and Systems, May 2010, pp. 253–256.
[7] Y. Lecun, L. Bottou, Y. Bengio, and P. Haffner, "Gradient-based learning applied to document recognition," Proc. IEEE, vol. 86, no. 11, pp. 2278–2324, Nov. 1998.
[8] G. Litjens et al., "A survey on deep learning in medical image analysis," Med. Image Anal., vol. 42, pp. 60–88, Dec. 2017.
[9] C. Chen, A. Seff, A. Kornhauser, and J. Xiao, "DeepDriving: Learning Affordance for Direct Perception in Autonomous Driving," presented at the Proceedings of the IEEE International Conference on Computer Vision, 2015, pp. 2722–2730.
[10] Y. Taigman, M. Yang, M. Ranzato, and L. Wolf, "DeepFace: Closing the Gap to Human-Level Performance in Face Verification," presented at the Proceedings of the IEEE Conference on Computer Vision and Pattern Recognition, 2014, pp. 1701–1708.
[11] A. Kamilaris and F. X. Prenafeta-Boldú, "Deep learning in agriculture: A survey," Comput. Electron. Agric., vol. 147, pp. 70–90, Apr. 2018.
[12] H. Li et al., "How MagNet: Machine Learning Framework for Modeling Power Magnetic Material Characteristics," IEEE Trans. Power Electron., vol. 38, no. 12, pp. 15829–15853, Dec. 2023.
[13] Mingxing Tian et al, "Neural Network Model for Magnetization Characteristics of Ferromagnetic Materials", IEEE Access, vol. 9, pp. 71236 - 71243, 2021.
[14] Can Ding, Yaolong Bai, Yinbo Ji, Pengcheng Ma, "Neural Network Modeling of Complex Hysteresis Loops in Ferromagnetic Materials", IEEJ Transactions on Electrical and Electronic Engineering, vol.20, no. 3, pp. 373-384, 2024.
[15] Sharareh Mirzaee, Kamran Sabahi, "Deep Artificial Neural Network Method for Magnetic Hysteresis Loop Prediction of Polyvinyl Alcohol@CoFe2O4 Nanocomposites", Transactions of the Indian Institute of Metals, vol. 77, pp. 2651–2657, 2024.